\documentclass[]{iopart}
\usepackage{graphicx}
\begin{document}
\title[Anisotropic magnetic properties and superzone gap formation in CeGe]{Anisotropic magnetic properties and superzone gap formation in CeGe single crystal}
\author{Pranab Kumar Das, Neeraj Kumar, R. Kulkarni, S. K. Dhar and A. Thamizhavel}
\address{Department of Condensed Matter Physics and Materials Science, Tata Institute of Fundamental Research, Homi Bhabha Road, Colaba, Mumbai 400 005, India.}
\ead{pkd@tifr.res.in}

\begin{abstract}
Single crystals of CeGe and its non-magnetic analogue LaGe have been grown by Czochralski method. CeGe compound crystallizes in the orthorhombic FeB-type crystal structure with the space group \textit{Pnma} (\#62). The anisotropic magnetic properties have been investigated on well oriented single crystals by measuring the magnetic susceptibility, electrical resistivity and heat capacity. It has been found that CeGe orders antiferromagnetically at 10.5~K. Both transport and magnetic studies have revealed large anisotropy reflecting the orthorhombic crystal structure. The magnetization data at 1.8~K reveals metamagnetic transitions along the [010] direction at 4.8 and 6.4~T and along [100] direction at a critical field of 10.7~T, while the magnetization along [001] direction increases linearly without any anomaly up to a field of 16~T. From the magnetic susceptibility and the magnetization measurements it has been found that [010] direction is the easy axis of magnetization. The electrical resistivity along the three crystallographic directions exhibited an upturn at $T_{\rm N}$ indicating the superzone gap formation below $T_{\rm N}$ in this compound. We have performed the crystalline electric field (CEF) analysis on the magnetic susceptibility and the heat capacity data and found that the ground state is doublet and the splitting energies from the ground state to the first and second excited doublet states were estimated to be 39 and 111~K, respectively.
\end{abstract}
\pacs{81.10.-h, 71.27.+a, 71.70.Ch, 75.10.Dg, 75.50.Ee}
\submitto{\JPCM}
\maketitle

\section {Introduction}
The magnetism in cerium based intermetallic compounds, where the $4f$ level lies in close proximity to the Fermi level, arises due to the competition between the Ruderman-Kittel-Kasuya-Yosida (RKKY) interaction and the Kondo effect. This competition leads to various diverse ground states like magnetic ordering, valence instability, heavy fermion nature etc.; hence Ce based intermetallic compounds have been widely investigated for several decades. In continuation of our research on simple binary Ce systems~\cite{Pranab}, we have investigated the anisotropic magnetic properties of orthorhombic CeGe. CeX compounds where X = Si and Ge crystallize in the FeB-type orthorhombic crystal structure with the space group \textit{Pnma} (\#62). CeSi has been investigated on a polycrystalline sample by Shaheen~\cite{Shaheen} and found to order antiferromagnetically at $T_{\rm N}$ = 5.6~K. From heat capacity results Shaheen concluded that the ground state is a quartet, which is unusual as the Ce atoms occupy the orthorhombic site symmetry and therefore a doublet ground state is expected. On the other hand, Noguchi~\textit{et al}.~\cite{Noguchi} have grown a single crystal of CeSi and found that the ground state is a doublet with a magnetic easy axis along [010] direction and a saturation moment of 1.75~$\mu_{\rm B}$/Ce. Polycrystalline CeGe has also been investigated about four decades ago, by Buschow~\cite{Buschow} and very recently by Marcano~\textit{et al}~\cite{Marcano1, Marcano2} and found to order antiferromagnetically with a N\'{e}el temperature $T_{\rm N}$ = 10.8~K. From the heat capacity studies Marcano~\textit{et al}~\cite{Marcano2} claimed that the Kondo interaction is present in CeGe. This is interesting as Ref.~\cite{Shaheen} does not report any evidence of Kondo interaction in the isostructural CeSi which has a lower unit cell volume compared to CeGe. Furthermore, Marcano~\textit{et al}~\cite{Marcano2} observed that in CeGe at $T_{\rm N}$, the electrical resistivity increased with decrease in temperature indicating the signature of the magnetic superzone gap formation in this compound. Due to the relatively low crystal symmetry of CeGe, we felt it was worthwhile to grow a single crystal of CeGe and explore its expected anisotropic magnetic behavior using the techniques of magnetization, heat capacity and electrical resistivity. A description of the anisotropic physical properties is presented in the following section together with the crystal field analysis of magnetic susceptibility and heat capacity.

\section{Experiment}

We adopted Czochralski pulling method to grow the single crystals of CeGe and non-magnetic reference compound LaGe in a tetra-arc furnace. High purity starting metals of Ce and La (3N pure) and Ge (5N pure) taken in the stoichiometric ratio with a little excess of Ge, totaling about 10 gm were melted in a tetra-arc furnace and crystals were then pulled at the rate of 10~mm/hr using tungsten rod as a seed.  The cleavage plane in CeGe is perpendicular to the pulling direction and the as grown crystals were brittle. A small piece of crystal was then subjected to powder x-ray diffraction (XRD), using a PANalytical x-ray diffractometer with monochromatic Cu-K$_{\rm \alpha}$ radiation, to check the phase purity. The orientation of the crystals was done by back reflection Laue method. The dc magnetic susceptibility and the magnetization measurements were performed in the temperature range 1.8-300~K using a superconducting quantum interference device (SQUID) and vibrating sample magnetometer (VSM). The electrical resistivity was measured down to 1.9~K in a home made set up. The heat capacity  was measured using a Quantum Design physical property measurement system (PPMS).

\section{Results}
\subsection{X-ray diffraction}

\begin{figure}
\begin{center}
\includegraphics[width=0.8\textwidth]{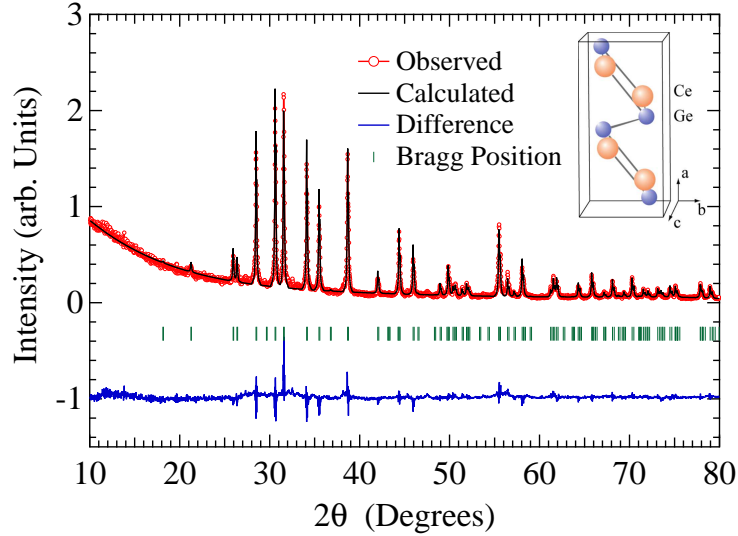}
\caption{\label{fig1} Powder x-ray diffraction pattern of  CeGe. The solid line through the experimental point is the Rietveld refinement pattern; the difference between the observed and the experimental pattern and the Bragg position are also shown. The inset shows the crystal structure of CeGe.}
\end{center}
\end{figure}

It has been reported by  Gokhale and Abbaschian~\cite{Gokhale} that CeGe forms peritectically at 1433~$^\circ$C  but increasing the Ge concentration slightly leads to a nearly congruent melting. Similar observation applies to LaGe. The powder x-ray diffraction pattern of our single crystals, grown from a slightly Ge rich melt shows the presence of single phase and no traces of secondary phases.  Figure~\ref{fig1} shows the XRD pattern for CeGe. A Rietveld analysis was performed using the FULLPROF software package~\cite{Fullprof}. A reasonably good fit of the experimental pattern confirms the orthorhombic space group $Pnma$ and the 
estimated lattice constants were $a~=~8.355(4)~\rm {\AA}$, $b~=~4.078(9)~\rm {\AA}$ and  $c~=~6.023(5)~\rm {\AA}$. From the Rietveld refinement we have found that both Ce and Ge occupy the $4c$ positions at (0.1788(3), 0.25, 0.6172(5)) and (0.0424(5), 0.25, 0.1366(11)), respectively. These values are in good agreement with the published data~\cite{Buschow, Marcano1, Hohnke}. The crystal structure of CeGe is shown in the inset of Figure~\ref{fig1}. The stoichiometry of the sample was further confirmed by the energy dispersive analysis by x-ray (EDAX) where the composition of the single crystal was analyzed at various locations. It was found that the sample is homogeneous maintaining the stoichiometry throughout the sample. The crystals were then oriented along the three principal crystallographic directions by means of Laue back reflection. Well defined diffraction spots corresponding to the orthorhombic symmetry pattern, further confirmed the good quality of the single crystals. The crystals were then cut along the principal crystallographic directions by means of a spark erosion cutting machine for the anisotropic physical property measurements.

\subsection{Magnetization}
\begin{figure}
\begin{center}
\includegraphics[width=0.8\textwidth]{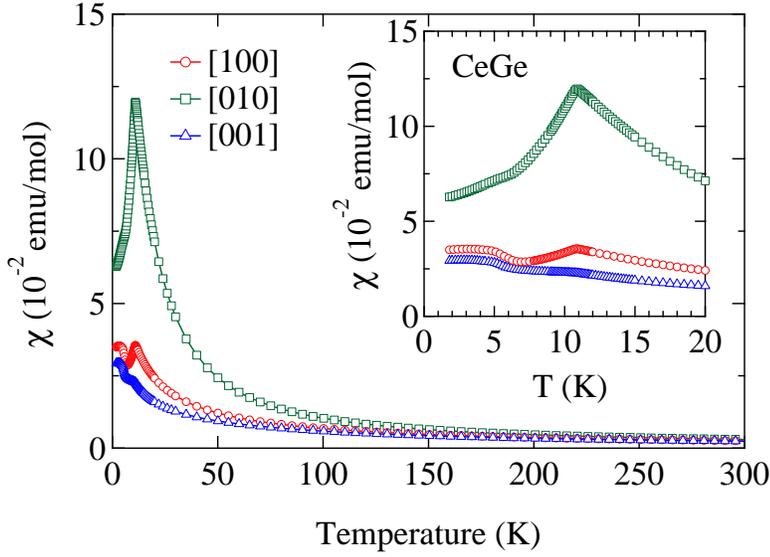}
\caption{\label{fig2} Temperature dependence of magnetic susceptibility of CeGe along [100], [010] and [001] directions. The inset shows the low temperature part of the magnetic susceptibility on an expanded scale.}
\end{center}
\end{figure}

The magnetic susceptibility of CeGe in the temperature range from 1.8  to 300~K in an applied magnetic field of 1~kOe for $H$ parallel to the three principal crystallographic directions viz., [100], [010] and [001] is shown in Fig.~\ref{fig2}. It is evident from the figure that the magnetic susceptibility is anisotropic both in the paramagnetic and in the magnetically ordered state. The 
inset shows the low temperature part of the magnetic susceptibility where the susceptibility exhibits a clear drop at 10.5~K for $H~\parallel$~[010] and a relatively smaller peak for $H~\parallel$~[100] direction while a small change of slope is observed for $H~\parallel$~[001] direction, indicating the onset of magnetic ordering at 10.5~K in this compound. The plots show that [010] is the magnetic easy axis which is further confirmed from the isothermal magnetization data to be discussed later. In an ideal antiferromagnet, the magnetic susceptibility ($\chi_{\parallel}$) along the easy axis should decrease to zero as the temperature is lowered to zero, while the susceptibility ($\chi_{\perp}$) is almost temperature independent along the hard axis. At low temperature, well below $T_{\rm N}$, the susceptibility data show an upturn indicating that the magnetic ordering in CeGe is not a simple collinear two sub-lattice antiferromagnetism. From the Curie-Weiss fitting in the temperature range (150 - 300~K), effective magnetic moment $\mu_{\rm eff}$ and the paramagnetic Curie temperature $\theta_{\rm p}$ were found to be 2.57~$\mu_{\rm B}$/Ce and -25~K, 2.61~$\mu_{\rm B}$/Ce and 16~K and 2.57~$\mu_{\rm B}$/Ce and -43~K, respectively for $H~\parallel~$[100], [010] and [001]. The experimental value of the effective moment $\mu_{\rm eff}$ is close to the free ion value of Ce$^{3+}$, 2.54~$\mu_{\rm B}$, indicating the local moment behaviour of Ce. The Weiss temperature  $\theta_{\rm p}$  is positive for $H~\parallel~$[010], while for the other two directions it is negative. The polycrystalline average of $\theta_{\rm p}$ is negative (-17~K) which is in conformity with the antiferromagnetic nature of the magnetic ordering.  

\begin{figure}
\begin{center}
\includegraphics[width=0.8\textwidth]{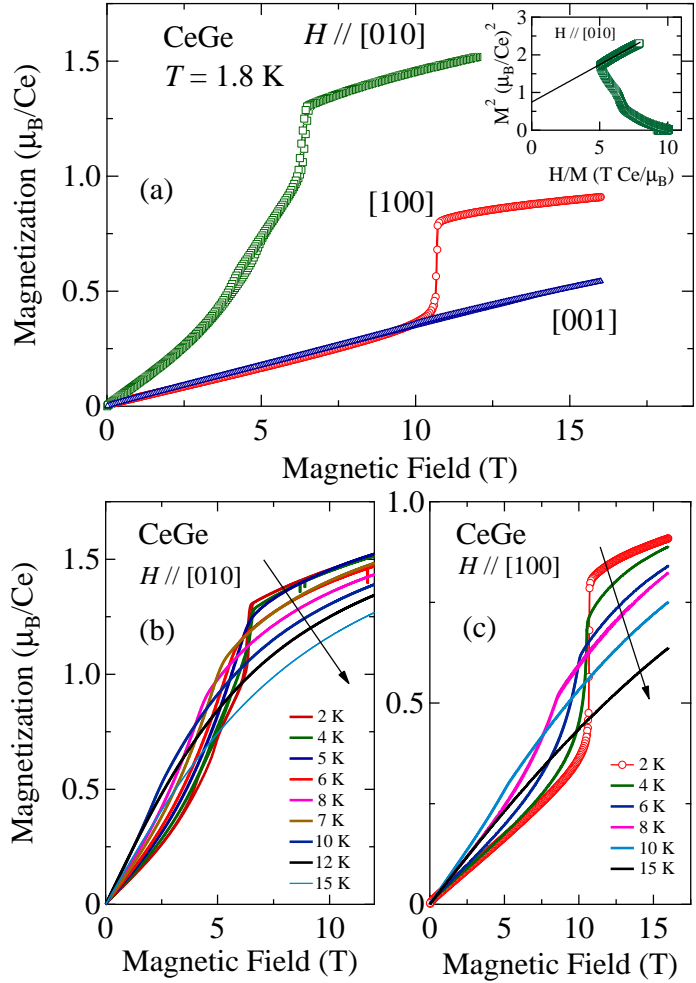}
\caption{\label{fig3}(a) Magnetization plots of CeGe along the three principal crystallographic directions measured at $T$~=~1.8~K. The inset shows the Arrot plot for $H~\parallel$~[010] magnetization data measured at $T$~=~1.8~K (b) Isothermal magnetization of CeGe measured at various fixed temperatures for $H~\parallel$~[010] and (c) for $H~\parallel$~[100] direction.}
\end{center}
\end{figure}

\begin{figure}
\begin{center}
\includegraphics[width=0.8\textwidth]{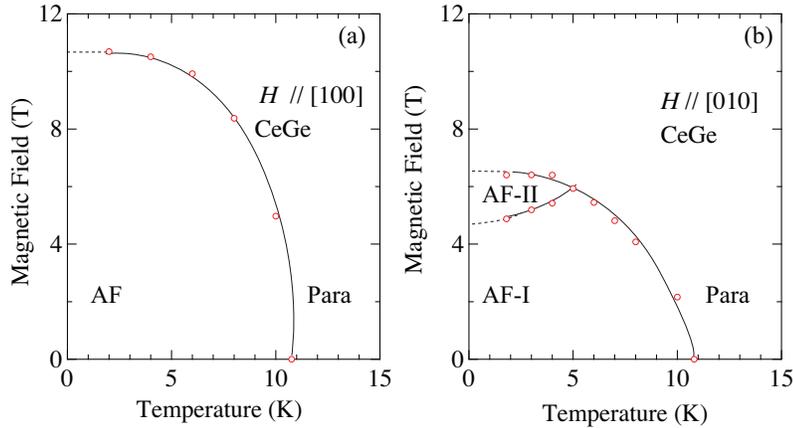}
\caption{\label{fig4}Magnetic phase diagram of CeGe for field parallel to (a) [100] and (b) [010] directions. The solid lines are guide to eyes.}
\end{center}
\end{figure}

The field dependence of magnetization $M(H)$ at a constant temperature $T$~=~1.8~K  is shown in Fig.~\ref{fig3}(a). For $H~\parallel~$[010] the magnetization is almost linear at low fields and exhibits an upward curvature as the field is increased. At a critical field of 4.8~T there is a small metamagnetic like transition indicating the first spin re-orientation. With further increase in the applied magnetic field, the magnetization undergoes a more prominent metamagnetic jump at 6.4~T and then increases gradually at higher fields.
The magnetization obtained at 12~T is about 1.5~$\mu_{\rm B}$/Ce which is less than the saturation moment of a free Ce$^{3+}$ ion($g_JJ\mu_{\rm B}$ = 2.14~$\mu_{\rm B}$).  An estimate of the ordered moment was done by using the Arrot plot~\cite{Arrot}.  The inset of Fig.~\ref{fig3}(a) shows the Arrot plot of the $H~\parallel$~[010] magnetization data measured at $T~=~1.8$~K.  The high field data was fitted to the expression $M^2$~=~$a\left(\frac{H}{M}\right) - b \epsilon$ where $a$ and $b$ are constants and $\epsilon$ = $(1-\frac{T_{\rm N}}{T})$.  From the fitting the ordered moment was estimated as 0.86~$\mu_{\rm B}$/Ce.  This corroborates with the previous neutron diffraction experiment on polycrystalline sample by Schobinger-Papamantellos \textit{et al}~\cite{Schobinger} where they estimated an ordered moment of 1.0~$\mu_{\rm B}$/Ce. They attributed the reduction in the moment value due to Kondo effect in this compound. Our resistivity measurement on single crystalline sample show clear signature of Kondo effect at high temperature (80 - 300~K) in this compound, which however vanishes at lower temperature to be discussed later. The magnetization along [100] and [001], on the other hand increases linearly with increase in the field almost uniformly up to a field of 10.6~T at which point, the magnetization along [100] exhibits a sharp metamagnetic transition and then shows a nearly saturation behaviour, while the magnetization along [001] continues to increase linearly without any anomaly. The data thus indicate that [001] direction is the hard axis of magnetization and [010] direction is the easy axis. The isothermal magnetization measurement was done along [100] and [010] direction at various temperatures. It is evident from  Fig.~\ref{fig3}(b) and (c), that with the increase in the temperature the metamagnetic transition shifts to lower fields and disappears at the ordering temperature. From the differential plots ($dM/dH$) of the isothermal magnetization, we constructed the magnetic phase diagrams which are shown in Fig.~\ref{fig4}(a) and (b). For $H~\parallel~$ [100] above $T_{\rm N}$ the compound is in the paramagnetic state, while it is in the antiferromagnetic state at lower temperatures.  At very low temperature for fields greater than 10.7~T where the magnetization exhibits a near saturation it is in the field induced ferromagnetic state. In the case of [010] direction, the small metamagnetic transition observed at 4.8~T increases to higher fields while the one at 6.4~T shifts to lower fields and they merge at around 6~T as the temperature of the isothermal magnetization is raised. At low temperature up to 5~K the system is in a complex antiferromagnetic state as defined by AF-II in the phase diagram.

\subsection{Electrical Resistivity}

\begin{figure}
\begin{center}
\includegraphics[width=0.8\textwidth]{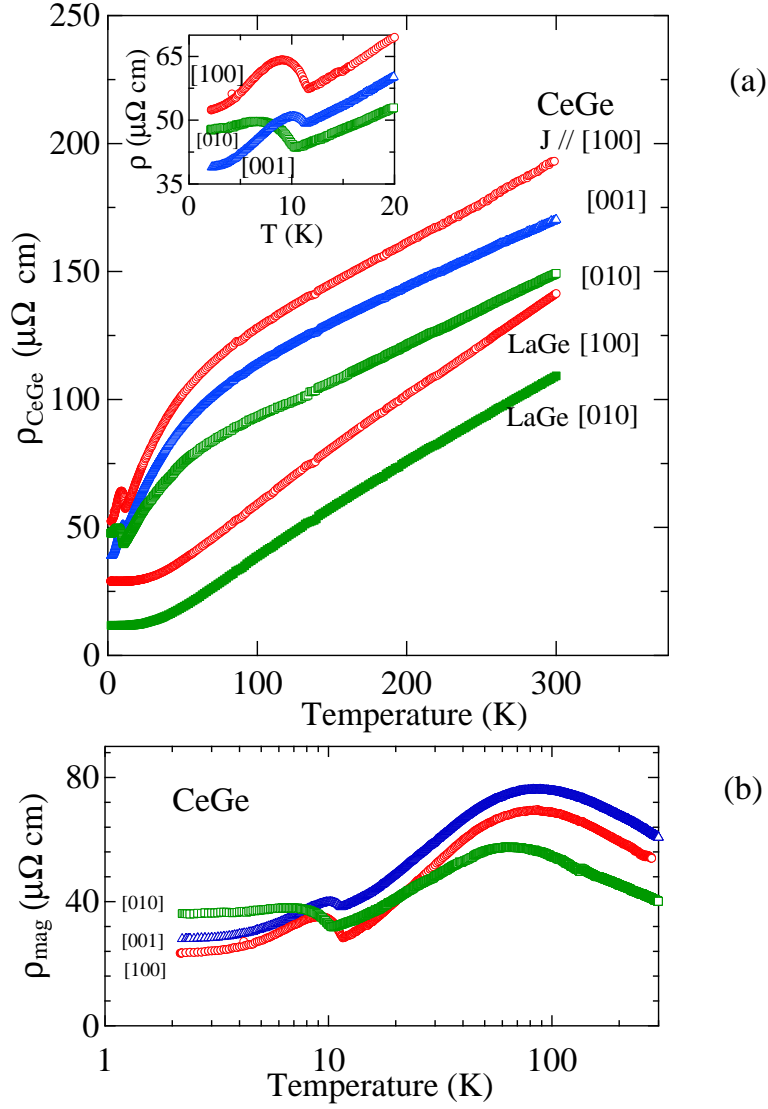}
\caption{\label{fig5}(a) Temperature dependence of electrical resistivity of CeGe and LaGe for current parallel to the three principal crystallographic direction.  The inset shows the low temperature resistivity of CeGe. (b) The magnetic part of the resistivity $\rho_{\rm mag}$ as a function of temperature in a semi-logarithmic scale.}
\end{center}
\end{figure}

Figure~\ref{fig5}(a) shows the electrical resistivity of CeGe and LaGe measured in the temperature range from 1.9 to 300~K for current parallel to the three principal crystallographic directions. The resistivity of LaGe is typical of a metallic system decreasing from 300~K to approximately 15~K, below which it is temperature independent with a residual resistivity of 11.7~$\mu \Omega \cdot$cm and 29.1~$\mu \Omega \cdot$cm for $J~\parallel$~[100] and [010] directions respectively thus indicating significant anisotropy. The resistivity of CeGe also decreases with the decrease in temperature and the residual resistivity (at $T$~=~2~K) of CeGe along the three principal direction is 52~$\mu \Omega \cdot$cm, 47.7~$\mu \Omega \cdot$cm and 38.96~$\mu \Omega \cdot$cm respectively for $J~\parallel$~[100], [010] and [001].
There is a broad hump at around 75~K for the current along the three principal directions.  Below 75~K the resistivity drops more rapidly down to the N\'{e}el temperature 10.5~K at which point the resistivity increases in all the three directions. For [100] and [001] the $\rho (T)$ shows a broad maximum centered around 9~K and then decreases as $T$ is decreased, while in the [010] direction the electrical resistivity levels off after passing through a broad maximum. This type of increase in the electrical resistivity at the magnetic ordering is attributed to the magnetic superzone gap effect, where the magnetic periodicity is greater than the lattice periodicity. If we consider the resistivity ratio $\rho(T = 2 K)/\rho(T = 15 K)$ we get the values 0.85, 1 and 0.71 for $J$ parallel to [100], [010] and [001] directions. This gives an indication that the superzone gap is larger in the $ab$-plane than along the [001] direction, which is in contrast to the case of heavy rare-earth metals where the superzone gap formation is along the hexagonal [001] direction~\cite{Elliott}. Our resistivity result is almost similar to the orthorhombic UCu$_2$Sn system~\cite{Takabatake} where the superzone gap effect is seen along all the three principal crystallographic directions. Figure~\ref{fig5}(b) shows the magnetic part of the electrical resistivity $\rho_{\rm mag}(T)$ obtained by subtracting the resistivity of LaGe from that of CeGe. The $\rho_{\rm mag}(T)$ is plotted on a logarithmic scale. A negative temperature coefficient of resistivity below 300~K with a broad peak at 80~K can be rationalized on the basis of incoherent Kondo scattering of the change carriers and the crystal field split levels. However, at low temperatures above $T_N$, we do not observe any evidence of Kondo effect in the resistivity data. As is well known, the Kondo temperature $T_K$ is proportional to the degeneracy of the $4f$ level and $T_K$ will decrease with the decrease in temperature due to the thermal variation of the occupied CEF levels~\cite{Yamada}.

\subsection{Heat Capacity}
\begin{figure}
\begin{center}
\includegraphics[width=0.8\textwidth]{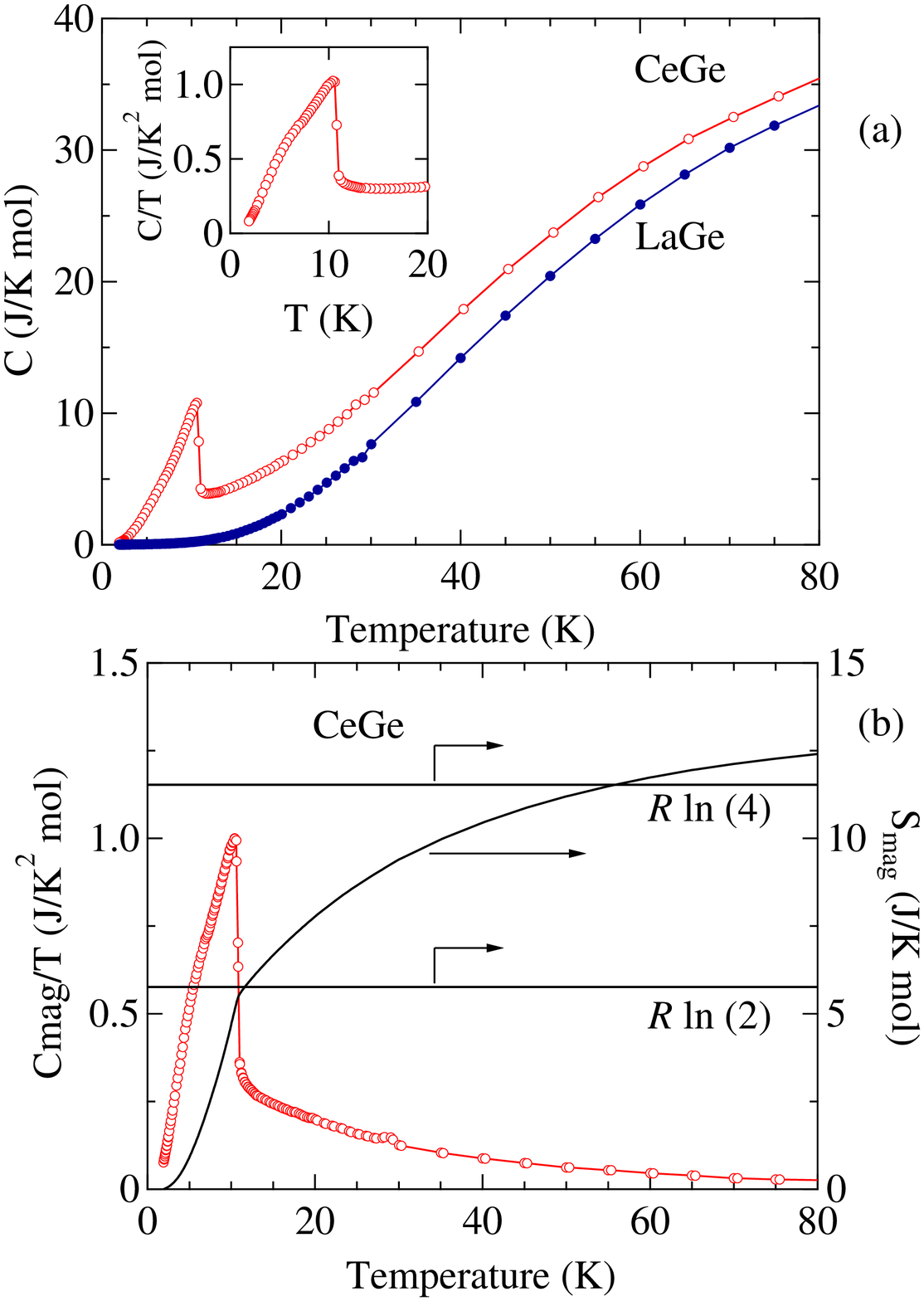}
\caption{\label{fig6}(a) Temperature dependence of the specific heat capacity in CeGe and LaGe. The inset shows the low temperature plot C/T vs. T of CeGe. (b)  C$_{\rm mag}$/T vs. T of CeGe. The calculated entropy is plotted in the right axis.}
\end{center}
\end{figure}

The temperature dependence of the specific heat of single crystalline CeGe and its non-magnetic reference compound LaGe measured in the range of 1.8 to 200~K is shown in the main panel of Fig.~\ref{fig6}(a). 
A very sharp transition is observed in the heat capacity of CeGe at 10.5~K confirming the bulk magnetic ordering. The heat capacity of CeGe is significantly larger than LaGe both in the ordered state and above $T_{\rm N}$.  The jump in the heat capacity of CeGe at the transition  amounts to 6.9~J~K$^{-1}\cdot$ mol$^{-1}$ which is substantially reduced compared to the mean field value of 12.5~J~K$^{-1}\cdot$ mol$^{-1}$ for spin $S$~=~1/2, but compares well with the value of 6.5~J~K$^{-1}\cdot$ mol$^{-1}$ observed by Marcano~\textit{et al}.\cite{Marcano2}. Usually, the reduction in the heat capacity from the mean field value is attributed to the Kondo effect. However, neither the magnetization data nor the low temperature electrical resistivity show any evidence of Kondo effect in CeGe and hence it may be concluded that the reduced jump in the heat capacity is due to the presence of background Schottky heat capacity arising from the thermal depopulation of the crystal field split levels as the temperature is decreased.  From the $C/T$ versus $T^2$ plot, by extrapolating the data below 4~K to $T$~=~0 we get the $\gamma$ ($C/T$ at $T=0$) value of 13~mJ/(K$^2\cdot$~mol)  while that for LaGe the Sommerfeld coefficient $\gamma$ is 5~mJ/(K$^2\cdot$ mol). The low value of $\gamma$ indicates that there is no mass enhancement in CeGe compound. The magnetic part of the heat capacity $C_{\rm mag}$ is estimated by subtracting the heat capacity of LaGe from that of CeGe. Fig.~\ref{fig6}(b) shows $C_{\rm mag}/T$ versus $T$ along with the entropy $S_{\rm mag}$ obtained by integrating the $C_{\rm mag}/T$ data. The entropy reaches $R$~ln~2 (= 5.76~J~K$^{-1}\cdot$ mol$^{-1}$) at 11.6~K in conformity with a doublet ground state. In Ref.~\cite{Marcano2} the corresponding temperature was 12.5~K. It may be noted that our choice of nonmagnetic reference compound LaGe used to estimate $C_{mag}$ of CeGe is most likely more suitable and appropriate than YNi used in Ref.~\cite{Marcano2}. Above $T_{\rm N}$ the entropy increases gradually and reaches $R$~ln~4 at around 56~K. This roughly gives the estimate of the first excited state of the crystal field split levels which is discussed in the next section.  
\begin{figure}
\begin{center}
\includegraphics[width=0.8\textwidth]{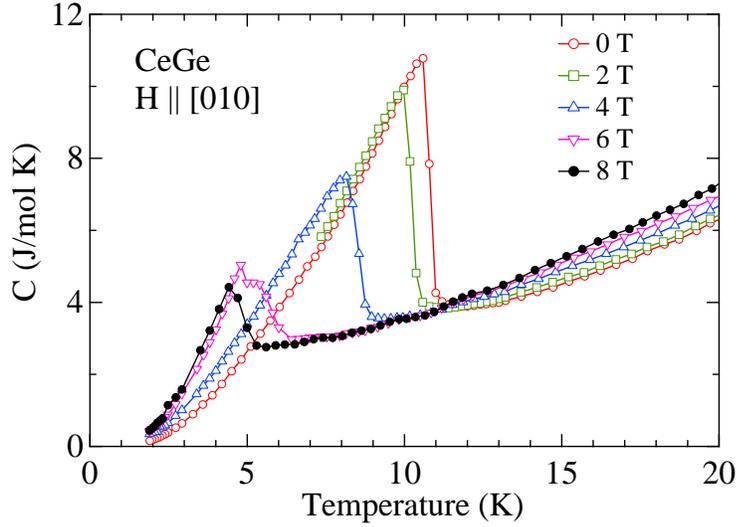}
\caption{\label{fig7}Heat capacity of CeGe in various applied magnetic fields with $H~\parallel$~[010] direction.}
\end{center}
\end{figure}
We have also measured the heat capacity of CeGe in various applied magnetic fields by applying the field parallel to [010] direction, which is the magnetic easy axis and the data are shown in Fig.~\ref{fig7}. It is evident from the figure that with the increase in the applied magnetic field, the heat capacity peak shifts to lower temperature. Such a behaviour is characteristic of antiferromagnetic compounds.

\section{Discussion}
From the magnetic susceptibility, electrical resistivity and heat capacity measurements it is obvious that CeGe undergoes an antiferromagnetic ordering at $T_{\rm N}$ = 10.5~K which is in agreement with the polycrystalline results by Marcano \textit{et al}.~\cite{Marcano1, Marcano2}. The measurements on single crystalline samples enabled us to identify the easy axis of magnetization as [010] direction. Furthermore, the ordered moment was estimated as 1.0~$\mu_{\rm B}$/Ce, from the isothermal magnetization along [010] direction at 2~K, by extrapolating the data above the second metamagnetic transition, to zero field as shown in Fig.~\ref{fig3}(a). Our results are in good agreement with the neutron diffraction results on polycrystalline sample~\cite{Schobinger}. The electrical resistivity of CeGe exhibits an unusual behaviour by undergoing an increase in the resistivity at the magnetic ordering temperature. Most of the cerium based compounds which order antiferromagnetically show a sharp decrease in the electrical resistivity due to the reduction in the spin disorder scattering. The increase in the resistivity at $T_{\rm N}$ is mainly attributed to the superzone gap formation in CeGe where the magnetic periodicity is larger than the lattice periodicity. At the antiferromagnetic superzone gap, some part of the Fermi surface will disappear and hence the resistivity increases. This effect is seen more prominently when the current is passed along the magnetic modulation direction.  Our resistivity data have shown an indication that the superzone gap is larger along the $ab$-plane than along the [001] direction.  

Next, we analyzed the magnetic susceptibility and heat capacity data using the CEF model. The Ce atom of the orthorhombic CeGe occupy the $4c$ Wyckoff position and hence possess the monoclinic site symmetry. For the monoclinic symmetry, the 6-fold degenerate levels of the $J$~=~5/2 multiplet will split into 3 doublets. In order to reduce the number of fitting parameters in the CEF analysis, we have used the CEF Hamiltonian for the orthorhombic site symmetry.  For Ce atom the Hamiltonian is given by,
\begin{equation}
\label{eqn2}
\fl
\mathcal{H}_{{\rm CEF}}=B_{2}^{0}O_{2}^{0}+B_{2}^{2}O_{2}^{2}+B_{4}^{0}O_{4}^{0}+B_{4}^{2}O_{4}^{2}+B_{4}^{4}O_{4}^{4} ,
\end{equation}
where $B_{\ell}^{m}$ and $O_{\ell}^{m}$ are the CEF parameters and
the Stevens operators, respectively~\cite{Hutchings,Stevens}. 

The magnetic susceptibility including the molecular field contribution $\lambda_{i}$ is given
by
\begin{equation}
\label{eqn3}
\chi^{-1}_{i} = \chi_{{\rm CEF}i}^{-1} - \lambda_{i}.
\end{equation}
\begin{figure}
\begin{center}
\includegraphics[width=0.8\textwidth]{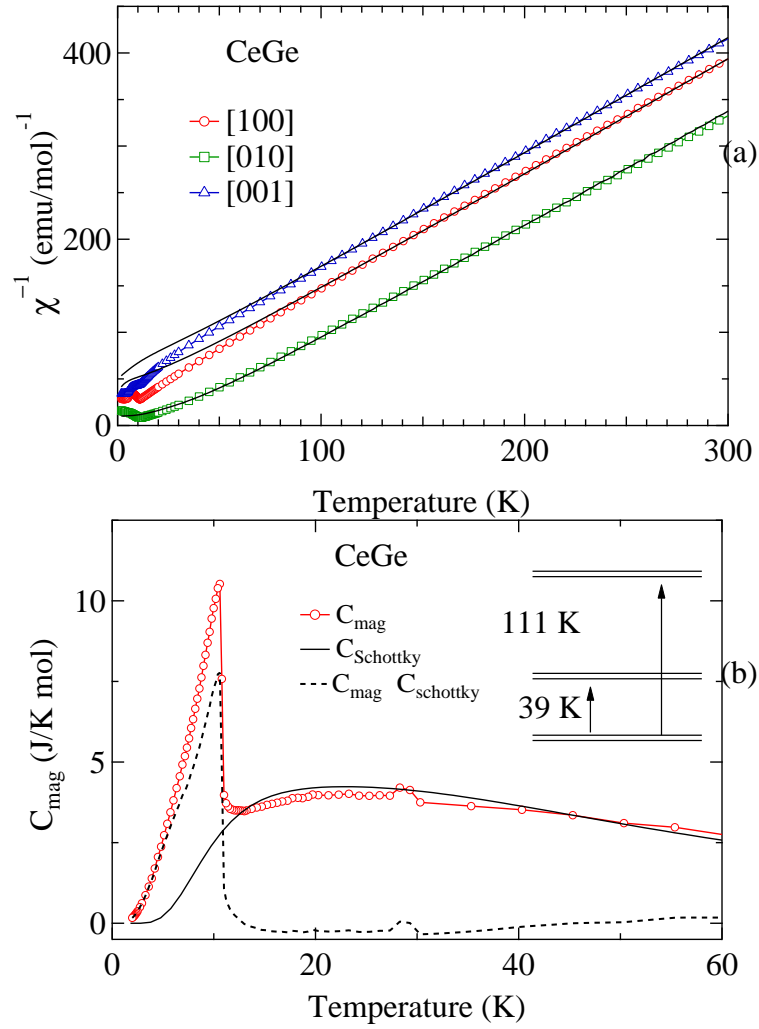}
\caption{\label{fig8} (a) Inverse magnetic susceptibility of CeCe. The solid line is based on the CEF calculation. The obtained energy levels are also shown (b) magnetic part of the specific heat capacity of CeGe. The solid line is calculated Schottky heat capacity. The dashed line indicates the magnetic part of the heat capacity with the Schottky heat capacity subtracted.}
\end{center}
\end{figure}
The expression for the magnetic susceptibility based on the CEF model is given in Ref.~\cite{Pranab}. We have calculated the inverse magnetic susceptibility for field along the three principal crystallographic directions based on the above CEF model. In these calculations, we have defined the [010] direction as the $z$-axis, the [100] direction as the $y$-axis and [001] direction as the $x$-axis while calculating the CEF susceptibility. The solid lines in Fig.~\ref{fig8}(a) are the calculated susceptibility with the unique value of the crystal field parameters as given in Table~\ref{table1}. The corresponding wavefunctions are also given in the table.

\begin{table}
\fl
\caption{\label{table1} CEF parameters, energy level schemes
and the corresponding wave functions for CeGe.}
\begin{tabular}{ccccccc} \hline
CEF parameters \\ \hline \\
& $B_{2}^{0}$~(K) & $B_2^2$~(K) & $B_{4}^{0}$~(K) & $B_4^2$~(K) & $B_{4}^{4}$~(K) & $\lambda_{i}$~(emu/mol)$^{-1}$ \\
& $3.0$ & $0.86$ & $0.01$ & $-0.48$ & $1.71$ & $\lambda_{[100]}$ = $-34$ \\
& & & & & & $\lambda_{[010]}$ = $61$     \\ 
& & & & & & $\lambda_{[001]}$ = $-50$ \\ \hline
$E$(K) & $\mid+5/2\rangle$ & $\mid+3/2\rangle$ & $\mid+1/2\rangle$ &  $\mid-1/2\rangle$ & $\mid-3/2\rangle$ & $\mid-5/2\rangle$ \\ \hline
111 & $0.834$ & 0 & -0.018 & 0 & $0.551$ & 0 \\
111 & 0 & $0.551$ & 0 & -0.018 & 0 & $0.834$ \\
39 & $0.011$ & $0.463$ & $-0.033$ & $0.837$ & $-0.018$ & $-0.288$ \\
39 & $0.288$ & $-0.018$ & $-0.837$ & $-0.033$ & $-0.463$ & $0.011$ \\
0 & $0.470$ & $0$ & $0.546$ & 0 & $-0.694$ & 0 \\
0 & 0 & $0.694$ & 0 & $-0.546$ & 0 & $-0.470$ \\ \hline
\end{tabular}
\end{table}
The negative value of the exchange field constant indicates the antiferromagnetic interaction between Ce moments. However, for $H~\parallel$~[010], the exchange field constant is positive indicating a ferromagnetic type of interaction of the moments along that direction. This is in accordance with the Weiss temperature, where a positive $\theta_{\rm p}$ was obtained from the Curie-Weiss fitting of the high temperature susceptibility data. By diagonalizing the Hamiltonian given in Eqn.~\ref{eqn2} using the estimated crystal field parameters we have obtained the crystal field split energies as $\Delta_1$~=~39~K and $\Delta_2$~=~111~K.

The magnetic part of the heat capacity of CeGe obtained after subtracting the heat capacity of LaGe, shown in Fig.~\ref{fig8} exhibits a broad peak in the paramagnetic state which is attributed to the Schottky type excitations between the CEF levels of the Ce$^{3+}$ ions. For a three level system (ground state plus two excited states), the Schottky heat capacity is given by the following expression
\begin{equation}
\label{eqn4}
\fl
C_{\rm Sch}= \left(\frac{R}{(k_{\rm B}T)^2} \frac{e^{(\Delta_1 + \Delta_2)/k_{\rm B}T}[-2 \Delta_1 \Delta_2 + \Delta_2^2 (1 + e^{\Delta_1/k_{\rm B}T}) + \Delta_1^2 (1 + e^{\Delta_2/k_{\rm B}T})]}{(e^{\Delta_1/k_{\rm B}T} + e^{\Delta_2 / k_{\rm B}T} + e^{(\Delta_1 + \Delta_2)/k_{\rm B}T})^2} \right)
\end{equation}
where $R$ is the gas constant and $\Delta_1$ and $\Delta_2$ are the crystal field split excited energy levels. The solid line in Fig.~\ref{fig8}(b) shows the calculated Schottky heat capacity as given by the Eqn.~\ref{eqn4}. The experimentally observed Schottky anomaly is satisfactorily reproduced by $\Delta_1$~=~39~K and $\Delta_2$~=~111~K obtained from the magnetic susceptibility data. It is to be mentioned here that $\Delta_1$ is close to the value at which the entropy reaches $R$~ln~4. The presence of a sizeable background heat capacity due to Schottky anomaly at the transition temperature $T_{\rm N}$ reduces the jump in the heat capacity at $T_{\rm N}$ from its mean field value of 12.5 J/(mol K) (for $S= 1/2$) to the observed value of 6.9~J/(mol K). It may be noted that if the Schottky contribution is subtracted from $C_{mag}$ then the entropy $R$~ln~2 due to the magnetic ordering alone will be attained at a temperature higher than 11.6~K as depicted in Fig.~\ref{fig6}(b). This indicates the presence of short range order above $T_{\rm N}$ which is not unusual. In Ref.~\cite{Marcano2} the authors did not take into account the Schottky contribution and the short range order above the $T_{\rm N}$. Therefore the validity of their analysis of the heat capacity data, based on the resonant level model, from which they inferred the presence of Kondo interaction at low temperature is doubtful. We have analyzed the magnetic part of the heat capacity by a model proposed by Blanco~\textit{et al}~\cite{Blanco}, where they have derived the expressions for the equal moment (EM) and the amplitude modulated (AM) magnetic structures from the jump in the heat capacity. For the case of equal moment structure, the jump in the heat capacity at the ordering temperature is given by,
\begin{equation}
\Delta C_{\rm EM} = 5 \frac{J(J+1)}{(2J^2 + 2J + 1)} R,
\end{equation}
and for the amplitude modulated system,
\begin{equation}
\Delta C_{\rm AM} = \frac{10}{3} \frac{J(J+1)}{(2J^2 + 2J + 1)} R,
\end{equation}
where $J$ is the total angular momentum and R is the gas constant. For spin S~=~1/2 system, the equal moment case $\Delta C_{\rm EM}$ amounts to 12.47~J/K$\cdot$mol and for the amplitude modulated case $\Delta C_{\rm AM}$ amounts to 8.31~J/K$\cdot$mol.  It is evident from the dashed line in Fig.~\ref{fig8}(b) that the jump in the Schottky subtracted magnetic part of the specific heat amounts to 8.26~J/K$\cdot$mol which suggests an amplitude modulated magnetic structure for CeGe, which is in conformity with the neutron diffraction results of polycrystalline CeGe. In order to understand further the magnetic properties exhibited by CeGe and to determine the magnetic structure, neutron diffraction experiments on the single crystalline sample are necessary which is planned for the future.

\section{Conclusion}

Single crystals of CeGe and LaGe were grown by Czochralski pulling method in a tetra-arc furnace. The transport and the magnetic measurements clearly indicate that CeGe orders antiferromagnetically at $T_{\rm N}$~=~10.5~K. A strong anisotropy is observed in the magnetization data reflecting the orthorhombic symmetry of the crystal. Field induced magnetic transitions were observed along the [100] and [010] directions while no such transition was observed along the [001] direction which is therefore the hard axis of magnetization. It is obvious from our measurements that [010] direction is the easy axis of magnetization. From the isothermal magnetization data measured at various fixed temperatures, we have constructed a tentative magnetic phase diagram for CeGe. The electrical resistivity showed an upturn at the magnetic ordering temperature there by indicating the superzone gap formation in CeGe along all the three principal crystallographic directions. We have found that the superzone gap is larger in the $ab$-plane than along the [001] direction. The electrical resistivity data show a possible presence of Kondo interaction at high temperatures which weakens and vanishes at low temperatures due to a reduced 4$f$-level degeneracy arising from the thermal depopulation of excited crystal electric field levels. The low value of the Sommerfeld coefficient and the entropy reaching $R$~ln~2 close to $T_{\rm N}$ confirms the localized nature of the Ce moments in this compound. The crystal field analysis on the magnetic susceptibility and the heat capacity has revealed the splitting energies as 39 and 111~K respectively for the first and the second excited states.  The heat capacity analysis clearly indicate the evidence of amplitude modulated magnetic structure of CeGe.

\end{document}